\documentclass{article}
\pdfoutput=1 
\usepackage{amsmath}
\usepackage{times}
\usepackage{graphicx}
\usepackage{float}
\usepackage{color}
\usepackage{multirow}
\usepackage[authoryear]{natbib}
\usepackage{rotating}
\usepackage{bbm}
\usepackage{latexsym}
\usepackage{amssymb}
\usepackage{multicol}

\usepackage{fancyhdr}
\usepackage{setspace} 

\begin{document}

\pagestyle{fancy}
\thispagestyle{empty}
\rhead{The Origin of Inference}

\textheight 23.4cm
\textwidth 14.65cm
\oddsidemargin 0.375in
\evensidemargin 0.375in
\topmargin  -0.55in
\headheight 15.0pt
\renewcommand{\baselinestretch}{2}
\interfootnotelinepenalty=10000
\renewcommand{\thesubsubsection}{\arabic{section}.\arabic{subsubsection}}
\newcommand{\myparagraph}[1]{\ \\{\em #1}.\ \ }
\newcommand{\citealtt}[1]{\citeauthor{#1},\citeyear{#1}}
\newcommand{\myycite}[1]{\citep{#1}}

\newcommand{\captionfonts}{\normalsize}

\makeatletter  
\long\def\@makecaption#1#2{%
  \vskip\abovecaptionskip
  \sbox\@tempboxa{{\captionfonts #1: #2}}%
  \ifdim \wd\@tempboxa >\hsize
    {\captionfonts #1: #2\par}
  \else
    \hbox to\hsize{\hfil\box\@tempboxa\hfil}%
  \fi
  \vskip\belowcaptionskip}
\makeatother   

\renewcommand{\thefootnote}{\normalsize \arabic{footnote}}

\hspace{13.9cm}

\vspace{20mm}
\begin{center}
{\bf \LARGE The Origin of Inference: \\ \vspace{5mm} \large Ediacaran Ecology and the Evolution of Bayesian Brains. }
\end{center}
\vspace{20mm}

{\large Michael G. Paulin}
\newline

{Department of Zoology, University of Otago, New Zealand.}\\

mike.paulin@otago.ac.nz

\vspace{10mm}

\vspace{20mm}

Running Head: The Origin of Inference.

%
DRAFT manuscript March 2015.  

\thispagestyle{empty}
%
\ \vspace{-0mm}


\pagebreak

\begin{center} {\bf Abstract} \end{center}
The evolution of spiking neurons and nervous systems in the late Ediacaran period  simultaneously with the evolution of carnivores around 550 million years ago can be explained by the need for accurately timed decisions under an imminent threat of being eaten. A simple model shows that threshold triggering devices, spiking neurons, are utility-maximizing decision-makers for the timing of escape reflexes given the sensory cues available to Ediacaran animals at the onset of carnivory.  Decisions are suboptimal for very weak stimuli, providing selection pressure for secondary processing of primary spike train data. A simple network can make approximately Bayes optimal decisions given stochastic spike trains. Decisions that are arbitrarily close to Bayes optimal can be obtained by enlarging this network.  A subnetwork that computes the Bayesian posterior density of the critical state variable - distance between predator and prey  - emerges as a core component of the decision-making mechanism. This is a neural analog of a Bayesian particle filter with cerebellar-like architecture. The model explains fundamental properties of neurons and nervous systems in modern animals and makes testable predictions.

\pagebreak

\section{Introduction} 
According to the Bayesian brain hypothesis, brains can compute Bayesian posterior probabilities from sense data and make Bayes-optimal decisions \citep{Kording2014, Pouget13, Lochmann11, Shams2010, Rao2010, Chater10, Wolpert2007, Doya07}. Much of the evidence for this is controversial \citep{hahn2014}. A major confounding factor is that in experimental behavioural economics it is often unclear whether subjects evaluate costs and benefits according to the intentions of the experimenter, even if they say they do.  Conversely, it is generally possible to reverse-engineer a Bayesian model to make any arbitrary behavior seem Bayes-optimal \citep{JonesLove2011, Bowers2012}.   

Predator-prey interactions between motile animals provide  useful model systems for clarifying this argument.  Direct energetic costs associated with an imminent threat of being eaten are very large relative to any other considerations. As a result, the short-term energetic consequence of a decision about whether and when to flee when a predator approaches is a natural measure of Bayesian utility in a natural ecosystem \citep{Feldman2010, Chittka2009}.  Nothing else matters in such confrontations.  Animals appear to be unambiguously Bayes-optimal decision-makers when their life is at stake \citep{Zylberberg2011,Neumeister2010,Schuster2010,Herberholz2012}. 

Nervous systems appear to have evolved in the late Ediacaran period about 550 million years ago (Ma), at the same time that animals first started eating each other. Animals  without neurons or nervous systems existed tens of millions of years earlier and became common after the Marinoan glaciation event which defines the start of the Ediacaran period at 632Ma.  The first animals fed by absorbing nutrients directly from seawater, via symbiosis with chemotrophic or phototrophic bacteria, or by suspension feeding.  Animals with all of these feeding modes exist today. By the late Ediacaran, near 550Ma, there were motile benthic algal grazers resembling modern placozoans and, probably, pelagic suspension feeders resembling modern ctenophores \citep{Monk2014a,ErwinBook2013}.  

For grazing or suspension-feeding animals, the net rate of energy intake is dependent on diffuse nutrient concentrations in algal mats or plankton clouds. These animals can use environmental cues that correlate with prey density to improve foraging efficiency and, if motile, to move into regions of higher energy density by following the cue gradients.  Such behaviors do not require a nervous system, as illustrated by taxes and trophisms of protists, plants, sponge larvae and placozoans in modern ecosystems.  Indeed, information about the location of individual prey has negative utility, because pursuing individual small organisms in water necessarily costs more than it is worth  \citep{Purcell1977}. 

The game changes when carnivores appear, for example if an animal evolves an enzyme that can digest animal protein. This carnivore can obtain large packets of energy in episodic encounters with other animals.  It then becomes worthwhile for grazers to withdraw from or avoid contact with other animals, and for carnivores to pursue them.

Pursuit could have been implemented by ciliated Precambrian metazoans lacking nervous systems, using gradient-tracking mechanisms inherited from their immediate ancestors.  Ciliated epithelial cells can both generate and sense forces, and could have served initially for contact detection and as actuators for pursuit and evasion. Ancestral ciliated epithelial cells have evolved into   receptor cells in the special sense organs of modern animals, capable of detecting a variety of specific sensory cues, including light, sound, electric fields, fluid movement in the environment and orientation and movement of the self in space \citep{Ludeman14,Frings2012, Cabej12, Fritzsch2004}.  These senses are vital for spatial orientation, perception and movement control in modern animals.  We  still use ciliated epithelia for actuation, in the lining of the lungs and in the ears, but forces for locomotion and body configuration changes are generated by muscles in most modern animal species.  Placozoans, ctenophores and some larval forms retain ciliary locomotion.  Fossil trace evidence for locomotion using muscles appears in the late Ediacaran, simultaneously with the emergence of carnivory \citep{Monk2014a}.

The very large energetic costs and benefits associated with contacting another animal would have provided strong selection pressure for the earliest carnivores and their prey to develop mechanisms for attack and evasion triggered by contact with another animal or, even better, by predicting contact before it happens.  

Fossil and molecular evidence indicates that the origin of neurons, nervous systems and muscles is causally related to the origin of of carnivory  \citep{Monk2014a}. This evidence is supported by mathematical models and sequential analysis of iterated predator-prey interactions in evolving populations.  The analysis shows that action potentials cannot have evolved unless there was an expected net benefit for expending a large amount of energy quickly at spike times. Conversely, when such costs and benefits exist, spiking neurons are almost certain to evolve if this is possible  \citep{Monk2014b,Monk2015b}.  

Molecular evidence shows that a proto-neuronal molecular genetic toolkit, including structural genes closely related to voltage-gated ion channels of spiking neurons, existed hundreds of millions of years before the first neurons evolved.  Close homologs of genes that were ultimately used to construct spiking neurons existed not only before neurons, but long before animals and even before multicellularity \citep{Li2015,Leys2015,Srivastava10,begemann2008,Maldonado04}.   Recent evidence  indicates that neurons and nervous systems evolved independently at least twice among Ediacaran metazoans \citep{Halanych2015, moroz2015, moroz2012}. Within the temporal resolution of the data, this simultaneous expensive innovation occurred at the same time that carnivores appeared in the fossil record.  This provides strong evidence that the appearance of carnivory in the late Ediacaran period was indeed the tipping point at which neurons switched from being unaffordable to being essential in those lineages \citep{Monk2014a}.

The escalation of biological armor and weaponry in association with the explosion of animal diversity in the early Cambrian period is well documented, and it is generally accepted that nervous systems and `intelligence' were part of the feedback loop that drove this event \citep{ErwinBook2013, dzik2007, Marshall06, McMenaminBook1990}.  But the general assumption appears to be that the earliest carnivores had neurons, which enabled them to detect and pursue prey, and that innovations such as shells, claws, spines, teeth etc. were a response to that threat.  The implication of the sequential analysis model is that neurons cannot have evolved before carnivory, but were necessary if they were possible as soon as carnivores appeared \citep{Monk2014b,Monk2015b}.   
 
In this paper, models and analysis are used to show that Bayesian decision theory not only provides a framework for predicting how a late Ediacaran grazer ought to behave when a predator approaches, but also predicts properties of underlying mechanisms necessary to implement these behaviors.  It will be shown that the mechanisms theoretically required for Bayes-optimal predator evasion in that ancient ocean correspond to fundamental design features of neurons and nervous systems of modern animals.

\color[rgb]{0,0,0}

\section{Optimal predator avoidance}
What should a defenseless, motile grazer do when a predator approaches?   There may be a large price to pay if it carries on regardless, grazing or engaging in some other activity that contributes to reproductive success, but there is also a price to pay if it interrupts what it is doing. There is not only the direct energetic cost of an avoidance maneuver, but also the opportunity cost of not feeding or flirting while fleeing.  At both extremes - too jittery or too confident in the face of danger - the grazer faces large costs.  In the intermediate regime it must be able to reduce expected losses by exerting itself briefly to lower the probability of being eaten when a threat is imminent. This intuition is supported by observations of modern animals, which move slowly most of the time with brief bursts of intense activity when predator and prey encounter each other \citep{Card2012,schusterfast2012,Zylberberg2011,domenici2011}. 

The situation can be analyzed using a simple model (Figure \ref{Cap:EDIACARAN}(a,b)).  In this model the capture rate, the probability of capture per unit time, depends on distance between predator and prey.  Studies of predator-prey interactions between motile animals in modern ecosystems indicate that this is the case.  The probability that a prey animal will survive a predatory strike is a sigmoidal function of distance between predator and prey when the strike is initiated, approaching zero when they are far apart and approaching one as they get closer \citep{Stewart2013}.  

Suppose that the prey is stationary at $x_0$ and the predator approaches at speed $v$, starting from $x=0$ where the probability of capture is negligible. By choice of units we can set $v=1$, so without loss of generality the predator is at $x=t$ at time $t$.  Assume that the capture rate at $x$ is $P(x)$, and that the prey gains energy at net rate $r$ while grazing and loses all of its stored energy $C$ if it is captured. 

In behavioural economics it is conventional to use dollars rather than Joules as the measure of utility, and the idea that being eaten has a finite cost is at least a little discomforting.  But evolution measures consequences at the population level.  Stored energy is a simple proxy for an individual's selective fitness, its capacity to contribute to the population, and in an evolutionary context it makes sense to assign a finite cost to the loss of an individual.

Suppose that the prey can perform some action making the probability of imminent capture negligible. Given the limited capacity of the earliest predators to detect and pursue prey, any small, random movement could have served this purpose at first.  This action resets the model to its initial state.  If the prey decides to perform such a maneuver at time $t$ then its net loss is

\begin{equation}\label{eq:LossTrial}
L(t) = - r t + \int_0^t{C P(x(t)) dt} + E,
\end{equation}
where $E$ is the energetic cost of the maneuver. This is minimized when 
\begin{equation}\label{eq:zeroNoiseOptimal}
P(x(t)) = r/C. 
\end{equation}

$P(x)$ and $x(t)$ are increasing functions, and so there is a unique solution for the critical decision time, when an escape maneuver will minimize the prey's expected loss, $t^* = x^{-1}(P^{-1}(r/C))$.

Equation (\ref{eq:zeroNoiseOptimal}) makes the unremarkable assertion that a prey animal should flee when the expected rate of energy loss due to being eaten equals the rate of energy gain from eating.  The specific form of the result is not important because the model is simplistic, although it does predict that grazers will linger longer before fleeing if they are feeding in a more nutritious patch (larger $r$) or if they are hungrier (smaller $C$), and this is true for modern animals \citep{Haag14, Chittka2009,Dill1997}.  The key result is that the optimal behavior is to continue grazing until the predator gets close, then flee rapidly and resume grazing.

It is common to say that timing is everything in games of agility, and this seems to be the case here. The critical time for the decision to flee is when the prey has just enough time to escape, which is when the time-to-collision is longer than the latency of the escape maneuver, with a certain fixed probability. But it is not necessary to have a clock to win this game.  Time-to-collision is a function of the state variable, $x(t)$, and the prey can determine the critical escape time, $t_c$, if it can determine when the predator reaches the critical state, $x_c = x(t_c)$.  The essential tool for winning this game is a mechanism for estimating this state variable.

\section{Cues and sensors} 

If the predator generates a signal that decays as a function of distance, $y= x_0-x$, between it and its prey, and the prey has a sensor that can measure this signal, then the prey can determine when the critical state is reached by determining when the signal from the predator reaches a critical level.  Escape behavior could be initiated by a threshold trigger mechanism. This does not require a nervous system.

Given that Ediacaran grazers already had force-sensing epithelial cells, the `natural selection' for a distance sensor would seem to be a mechanosensory hair cell that detected water displacement from an approaching predator. Indeed, displacement-sensing mechanosensory hair cells of the lateral-line system provide the crucial timing cues that modern fish use to trigger fast escape maneuvers during predatory strikes \citep{Nissanov89,Eaton77}. But there are two problems with this hypothesis.  The first is that the earliest predators, by assumption, moved using cilia.  They must have moved very slowly with very low Reynolds numbers, and would not have generated water displacements to betray their approach before contact \citep{Purcell1977}.  The second problem is that water displacement measurements are confounded by other sources: Waves, currents, turbulence and the motion of the prey itself.  While it is possible to isolate signals from a particular source in a noisy environment, this requires sophisticated signal processing, and these hypothetical Ediacaran ancestors did not have a nervous system at the time. 

Bioelectric fields provide a feasible solution.  Animals in seawater are surrounded by a dipole-like electric field due to osmoregulatory ion flux across epithelial tissues \citep{Bedore13,Kalmijn74}.  Ediacaran metazoans moving via cilia would have carried a standing electric field along with them.  This self-field is stationary in an animal's own reference frame, and the only localized electric fields in its environment are caused by other animals. Electroreceptor cells in modern fishes are modified hair cells, with accessory structures that focus electric fields, rather than mechanical forces, onto the transduction gates \citep{Fishman94,Fessard74}.  

Mechanical stimulation is ambiguous, even at contact, while localized electric fields in the Ediacaran ocean would have been unambiguous beacons attached to each animal.  Electric dipole field strength falls with the cube of distance to the source, so as a dipole gets very close the spatial gradient of the field becomes very strong.  This means that initially, when predators and prey are very slow-moving and the critical distance is small, the field would be strong and changing rapidly at the critical distance.   This would make it easy for a threshold trigger to evolve.  The rapid falloff in stimulus intensity with distance would mean that bioelectric field sensors could retain the spatial specificity of local contact forces, allowing the trigger to activate reflexes that cause movement away from the threat.  

Multiple cues could have been used to detect contact, but receptors responding to electric field cues should quickly take over the role of predator detection before contact, because this cue does not require a nervous system to interpret what it means, or what a grazer should do in response to it. A receptor responding to an electrical cue indicates that another animal is approaching the receptor, and if the field reaches a critical strength it is necessary to withdraw in the opposite direction.

Among all available sensory cues, the bioelectric field of an approaching animal seems to be uniquely capable of indicating the distance and direction to another animal in close proximity, before it makes contact, without interference from other causes.  It would have been possible for a motile animal with simple threshold triggers, but without a nervous system and without any capacity to perform causal inference, to optimally time its withdrawal from an approaching predator by activating a withdrawal reflex when the electric field strength reached a critical level.

\section{The effect of sensor noise}

Physical mechanisms can't measure signals without error, so a real threshold trigger must exhibit some jitter around the optimal escape time.  The jitter will depend on the signal-to-noise ratio (SNR) in the sensor (Figure 1c).  The situation is easier to understand in the discrete-time case, in which the sensor makes a sequence of measurements, $u_k = u(y(t_k))$, and triggers if $u_k > u_c$.  We can write $u_k = s_k + v_k$ where $s_k$ is the signal and $v_k$ is the noise in the $k$th sensor measurement.  The sensor will trigger if $v_k>u_c-s_k$.  The probability that this will occur is the integral of the upper tail of the noise density, $h(u)$,

\begin{equation}
p_k = p(\textrm{ spike } |  \textrm{ predator at } x_k ) = \int_{u_c-s_k(x_k)}^{\infty}{h(u) du}.
\end{equation}

\begin{figure}[h]\label{fig:EDIACARAN}
\centering
\includegraphics[width = 0.5 \textwidth]{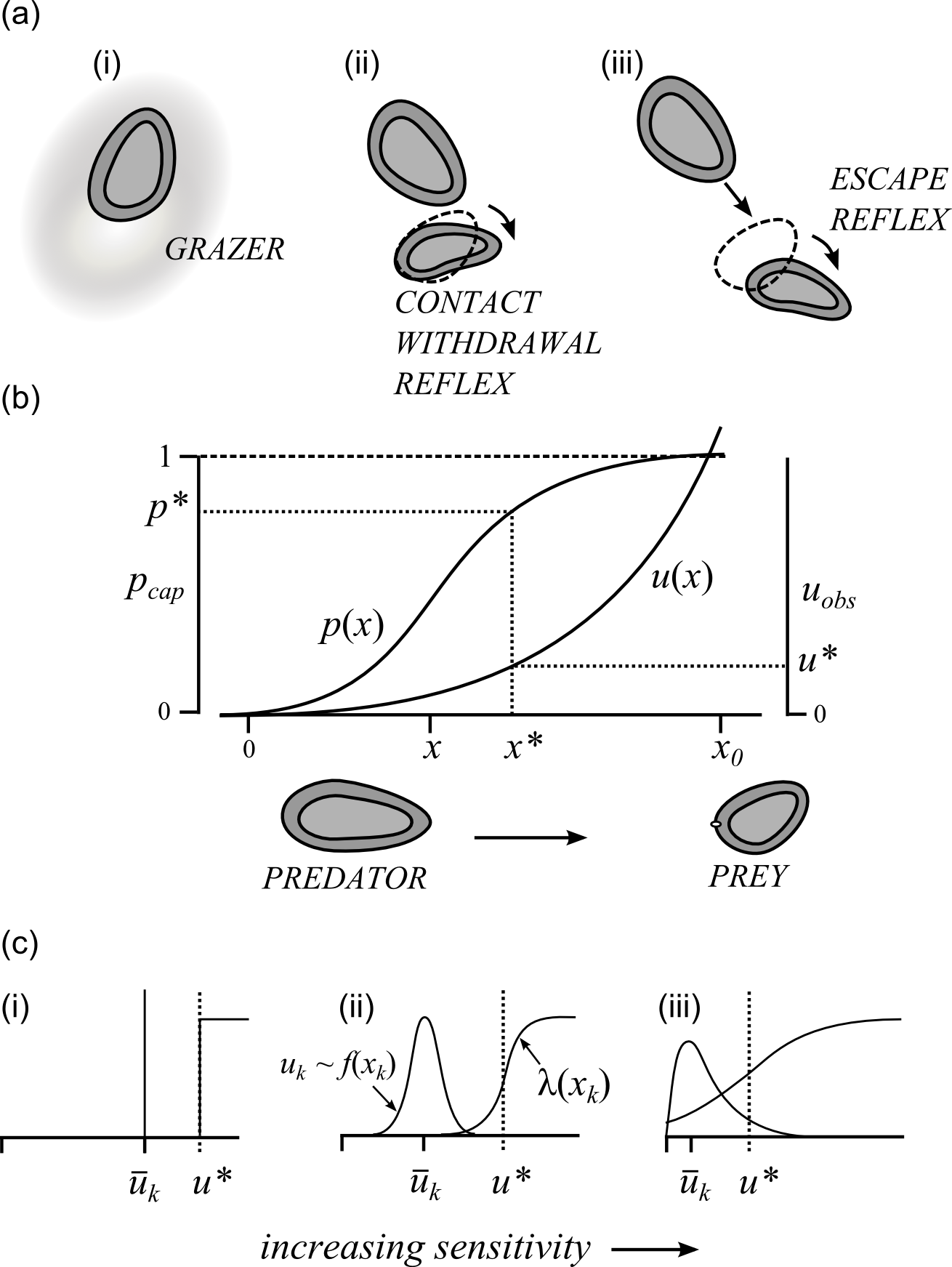}

\caption{ \label{Cap:EDIACARAN} (a) At the onset of carnivory, an Ediacaran benthic algal grazer (i) comes under selection pressure to withdraw from contact (ii) and then to flee before contact (iii).  (b) Probability of capture as a function of distance, $p(x)$, rises sigmoidally from negligible to certainty as the carnivore approaches.  A sensory cue, $u(x)$, increases as a power law function of distance, becoming very large with a very large gradient at close range.  The optimal decision is to flee  when the probability of capture reaches a critical level, $p^*$, corresponding to a critical sensory stimulus intensity, $u^*$. (c) In the noise-free limit, a threshold trigger is optimal (i).  If the trigger is noisy, the Bayes optimal decision is to wait for a sensory spike that pushes a critical mass of posterior probability above a threshold (ii).  At low signal-to-noise ratios, sensory spikes will occur spontaneously in the absence of a threat, because the tail of the measurement distribution is above the sensory firing threshold even when the stimulus is absent (iii).  }
\end{figure}

In the low-noise limit, $u_k = s_k$ is a deterministic function of the distance to the predator, and the prey flees at the first sample time after the predator reaches the critical point.  However, when there is noise the probability of a spike is a function of the distance to the predator.  If the SNR is high then this probability is near zero until the predator is almost at the critical point, and near 1 if the predator is only a little closer than the critical point.  On the other hand if the SNR is low then the firing probability changes gradually as the predator approaches. In the low SNR limit the sensor triggers randomly, independently of the location of the predator (Figure \ref{Cap:EDIACARAN}).

In discrete time the sensor firing sequence is a Bernoulli process with spiking probability $p_k$ in each sample period, increasing as the predator approaches.  In the continuous-time limit this becomes a Poisson process, with spiking intensity $\lambda(t)$ increasing as the predator approaches.  For high SNR the Poisson intensity is very low until the predator is near the critical point and increases rapidly as it gets closer.  The sensor is unlikely to trigger until the predator is nearly at the critical point, but very likely to trigger if it gets closer.  For lower SNR the Poisson intensity increases more gradually as the predator approaches and in the limit the firing pattern becomes random and independent of the predator's location.  

Because noise-free operation is physically impossible a real spiking sensor cannot trigger an optimal escape maneuver, but it can be near-optimal if the signal-to-noise level is high. Sharks, which have highly evolved electrosensory systems, can detect bioelectric fields of small animals in seawater several tens of centimeters away \citep{Bullock2005,Collin04,Kalmijn82}.  The critical distance for the earliest carnivores may have been in the order of a millimeter or even closer. At this range, because of the inverse cube law, the field would have been about a million times stronger than those detectable by shark electroreceptors today.  Thus a simple threshold trigger responding to electric fields may have provided `touch at a distance', allowing grazers to withdraw just before contact occurred. 

Evolution would select carnivores for agility and stealth, for mechanisms that increase the conditional probability of successful strikes at greater distances, and for weakening and disguising cues that betray them to prey.  This would push the critical distance further away and make the sensory cue(s) much weaker at the optimal decision time.  

Sensor noise levels are constrained by physics, ultimately by thermal noise in molecular mechanisms.  If predators can become so effective that the critical sensor level $u_c$ drops into the thermal noise floor it would seem that the game is up for fleeing as an anti-predation strategy.  At this point the `optimal' spiking sensor is spontaneously active when there are no predators around, and it fires only slightly more often if a predator approaches.  Costly escape maneuvers would be triggered at random, just a little more often when there is imminent danger of being eaten.  At that point it would seem that animals must retreat into shells and burrows, poking appendages out when conditions are less favorable for predators, or develop spines or toxins to avoid predation.  Indeed many animals chose such strategies, or combinations of them, as the predator-prey arms race exploded at the onset of the Cambrian period 543 million years ago \citep{ErwinBook2013,dzik2007, Marshall06}. 

\section{Principles of inference}

As the signal gets weaker relative to the noise, spikes from predator-detecting neurons are no longer confident assertions that a critical state, $x_c$, exists at that instant. Instead, spike trains become random samples from a Poisson process parameterized by the state, whatever the state happens to be at the time. The sensory signal, $u(t)$, may be all but invisible in the noise, with not enough energy to reliably trigger an escape maneuver, but stochastic spike trains carry information about distance to the predator in energetic packets.   These clearly-detectable pulses could provide data for a mechanism to infer the hidden state that parameterizes their statistical distribution.  

When information about the state is available only via noisy observations, the optimal decision is to minimize the conditional expected loss given those observations.  This is called Bayes optimality \citep{Berger2005}.  The difference between optimality and Bayes optimality is that `optimal' means the best that you can do given the state you are in, whereas `Bayes optimal' means the best that you can do given the information you have about the state  you are in.  Bayes-optimal is optimal if you have perfect knowledge of the world.

The conditional expected loss for escaping at time $t$ given observations $z(t)$ up to time $t$ is obtained by replacing the unknown location of the predator $x$ in equation \ref{eq:LossTrial} by its conditional probability distribution given the observations, $f(x|z)$, and integrating this over all possible locations,

\begin{equation}\label{eq:LossCondition}
\mathcal{L}(t) = - r t + \int_0^t{\int_0^{x_1}{C P(x) f(x|z(t)) dx} dt} + E.
\end{equation}

To evaluate loss in terms of expected reproductive success, the sigmoidal single-trial capture probability function for an individual, $P(x)$ in equation \ref{eq:LossTrial}, is replaced by the probability of lineage extinction.  This is a step function, 0 when $x<x_c$ and 1 when $x>x_c$ \citep{Monk2014b,Monk2015b}.  Differentiating $\mathcal{L}(t)$ and setting the result to 0 shows that the expected loss is minimized if the prey decides to flee when 

\begin{equation}\label{eq:LossCondition}
p_{crit} = Pr(\textrm{predator closer than } x_c | z(t)) = \int_{x_c}^{x_1}{f(x|z) dx}  = r/C.
\end{equation}

The optimal escape time is when the conditional probability that the predator is inside the critical region, given sense data obtained up to that time, exceeds a critical threshold.  

The conditional probability density function of some variable(s) of interest, given observations whose probability distribution depends on those variables, is known as the Bayesian posterior density \citep{Berger2005}.  Bayes-optimal decisions generally depend on the posterior density of states that affect the consequences of those decisions. In the current model, the consequences of a decision to flee depend on how far away the predator is at the time, and the Bayes optimal decision depends on the posterior density of that distance given spike trains whose statistical distribution depends on the distance.

The posterior density can be obtained from observations and the known conditional probability density of observations given the state by applying Bayes rule,

\begin{equation}
f(x|z) = \frac{f(z|x)f(x)}{p(z)}.
\end{equation}

It is less than 250 years since Laplace discovered this rule and showed how to apply it for causal inference \citep{Stigler86,Stigler1980}.  Although it has a very simple form, it can be  difficult to apply except in certain special cases, such as single-trial binary choice problems, or Kalman filtering for linear-Gaussian dynamical systems.  It's only within about the last 25 years that sufficiently powerful computers and algorithms have been developed to make Bayesian inference practical much beyond these special cases.  

It seems unlikely that brainless slug-like animals, being gradually culled from Ediacaran ecosystems as their once-effective escape trigger mechanisms became increasingly unreliable because predation pressure resulted in declining signal-to-noise ratios, could have pre-empted Laplace by well over half a billion years and saved themselves by discovering how to implement Bayes rule to infer the intensity of a Poisson process from samples in real time.

\section{Real-time inference from Poisson samples}\label{SECTION:Inference}
Inferring the probability that the distance to the predator is less than some value $u_c$ is equivalent to inferring the probability that the Poisson intensity  of the sensor is greater than some value $\lambda_c$.  So there is no loss of generality in considering how to infer the posterior density of $\lambda$ rather than of the distance.  Note that intensity, which is the parameter of a Poisson process at an instant, is distinct from rate, which is a property of a sample drawn from the process.  The rate of a sample is the number of spikes divided by its duration.  In general, the expected rate of a sample from a Poisson process is the average intensity over the duration of the sample \citep{Brillinger01,Landolt78}.  In the special case of a homogeneous Poisson process, in which the intensity is constant, intensity is the expected value of rate.

In a Poisson process with intensity $\lambda$, the probability that no spikes occur in an interval of duration $t$  is $e^{-\lambda t}$, and the probability that at least one spike occurs in this interval is $1 - e^{-\lambda t}$.  The probability that at least one spike occurs before time $r<t$ and no further spikes occur until $t$ is
\begin{equation}\label{eq:biased_likelihood}
e^{-\lambda \left( t - r \right)}\left( 1 - e^{-\lambda r} \right) = e^{-\lambda t} \left( e^{\lambda r} - 1\right) .
\end{equation}
This probability is the likelihood for $\lambda$ at time $t$ if at least one spike was observed in the interval $[0,r)$ and no further spikes have been observed.  

The likelihood goes to zero as $r$ goes to zero.  This means that the observer must only observe presence or absence of events in a finite time window, not the precise timing of events. The likelihood is unbiased if the spike occurs at a random time in the  window.

Unless the observer has some independent way to synchronize clocks with the sensor, they can only measure the time $t_s$ since a spike, not the time $t$ since the start of a time-window in which the sensor randomly emitted a spike.  The expected difference between these two times is the mean waiting time for an event in a Poisson process with parameter $\lambda$, given that an event has occurred in $[0,r)$. This is a function of $\lambda$ and $r$, 
\begin{equation}\label{eq:refractory_correction}
\Delta t  = 1/\lambda - r/(e^{\lambda r} - 1).
\end{equation}

Combining this with (\ref{eq:biased_likelihood}) gives an expression for the likelihood of $\lambda$ at time $t$, where $t$ is no longer the time since the unknown initial time $0$, but the known time since the most recent spike, under the assumption that no more than 1 spike occurs in an interval of duration $r$,
\begin{equation}\label{eq:pspike}
L \left( \lambda,  t , r\right) =e^{\lambda \Delta t}e^{-\lambda t} \left( e^{\lambda r} - 1\right).
\end{equation}

The assumption that at most one spike occurs in any interval of duration $r$ means that there must be some censoring process, either a refractory mechanism in the spike generator to prevent a second spike occurring, or in the observing mechanism to ignore a second spike if one does occur within $r$.  The fact that spikes are energetically expensive would seem to favor the first option.   

Now suppose that the prior probability density of $\lambda$, $f(\lambda, t_j)$, is known at the $k$th spike time.  Then the posterior density at time $t_j + t$ in the subsequent inter-spike interval is given by Bayes rule, which for this inference problem takes the form
\begin{align}\label{eq:BayesRule}
f\left(\lambda, t_j+t\right) &= \frac{e^{\lambda\Delta t}\left( e^{\lambda r}-1\right)e^{-\lambda t}  f\left( \lambda, t_j \right)}{\int_0^\infty \!  e^{\lambda \Delta t}\left( e^{\lambda r}-1\right)e^{-\lambda t}   f \left(\lambda, t_j \right) \textrm{d}\lambda}.
\end{align}
An initial prior, $f(\lambda, 0)$,  is chosen and at each spike time $t_j$ the posterior at that time, $f(\lambda, t_j)$, becomes the prior for inference during the next inter-spike interval. 

Equation \ref{eq:BayesRule} closely resembles the standard form of Bayes rule applied for sequential inference \citep{Doucet01}.  It differs from the standard form in that the data are not varying quantities observed at regular time intervals, but fixed pulses observed at varying time intervals.  This formula computes the posterior density at all times, not just at the observation times. 

The shape of the posterior density is specified by the numerator in equation \ref{eq:BayesRule}, while the denominator merely adjusts the scale.  The numerator changes smoothly during interspike intervals, falling at an exponentially faster rate for higher values of $\lambda$. As a result, probability mass flows continuously towards lower values of $\lambda$ during interspike intervals, and the shape of the posterior density becomes more exponential-like.  At spike times, when $t = t_{j+1}-t_j$,  $j$ is replaced with $(j+1)$,  $t$ is reset to zero and the post-spike numerator is formed by multiplying the pre-spike posterior by $e^{\lambda\Delta t}\left( e^{\lambda r}-1\right)$.  This factor is an exponentially increasing function of $\lambda$, so its effect is to shunt probability mass discontinuously towards larger values of $\lambda$ at spike times, and the shape of the posterior density becomes more Gaussian-like.

\section{Approximate Bayes-optimality with one neuron }
Because the posterior probability that the predator is inside the critical zone can increase only at spike times, the Bayes-optimal decision for a grazer must be to flee at a spike time.  But which one? 

The critical escape time is easy to determine numerically, by evaluating equation \ref{eq:BayesRule} sequentially at the spike times, computing the integral in the tail of the density above the critical intensity, $\lambda_c$, and stopping when this is above the critical probability, $p_{crit}$.  However, our hypothetical ancestor did not have a computer or even a nervous system at this time. It had only recently evolved spiking neurons.  

There is a simple way to estimate the critical threshold-crossing time without a digital computer, based on the fact that the Poisson intensity is the expected average rate of a spike train. This means that the average spike rate is expected to reach the critical level $\lambda_c$ when the predator reaches the critical distance $x_c$.  Longer intervals give less-variable estimates of $\lambda$, and threshold-crossing times closer to the optimal time. 

Average spike rate can be estimated using a low-pass filter. The simplest possibility is a first-order low pass filter, given by 
\begin{equation}\label{eq:avgrate}
\frac{dv}{dt} + v/\tau = w \sum{\delta(t-t_j)}
\end{equation}
where spikes occur at times $t_j$, $w$ is the spike weight or filter input gain, and $\tau$ is the filter time-constant.  

Under the simplifying assumptions that the Poisson intensity is large relative to the filter time constant (many spikes are expected to occur over the integration time of the filter, $\tau$) and that it changes only very slowly  relative to the filter time constant (the predator is approaching very slowly, making negligible progress over time $\tau$), the right-hand side of \ref{eq:avgrate}  can be approximated as a constant, $w \lambda$, representing the input as a steady flow rather than as a steady series of pulses. The solution is then
\begin{equation}\label{eq:leakyintegratesoln}
v(t) = w \lambda ( 1 - e^{-t/\tau}).
\end{equation}
This gets arbitrarily close to $w \lambda$ as $t$ gets large, and will go above $w \lambda_c$ if the predator is closer than the critical distance, $u_c$.  The predator can be detected at $u_c$ by setting a threshold trigger at the critical level, $w \lambda_c$.  

Equation \ref{eq:avgrate} is the standard leaky integrate-and-fire (LIF) model of a neuron, where $v(t)$ is membrane voltage referenced to resting potential \citep{burkitt06}. Synaptic currents in real neurons are proportional to membrane voltage referenced to the reversal potential of the synaptic channels, which is not generally the resting potential. This detail makes the solution (\ref{eq:leakyintegratesoln}) slightly different in detail when a LIF model is used rather than a biophysically realistic model, but does not change its general form and does not affect the following qualitative argument. 

The solution (\ref{eq:leakyintegratesoln}) indicates that the Bayes-optimal escape time can be computed by using a secondary neuron to smooth the sensory spike trains and trigger when the average firing rate reaches a critical level.  Unfortunately, this result is only asymptotically Bayes-optimal. It is exact in the uninteresting limiting case in which the predator is not moving and the prey can take an arbitrarily long time to decide.  In relevant cases the predator moves and the prey is under time pressure to make a decision.

\begin{figure}[h]\label{fig:LIFTRIGGER}
\centering
\includegraphics[width = 0.5 \textwidth]{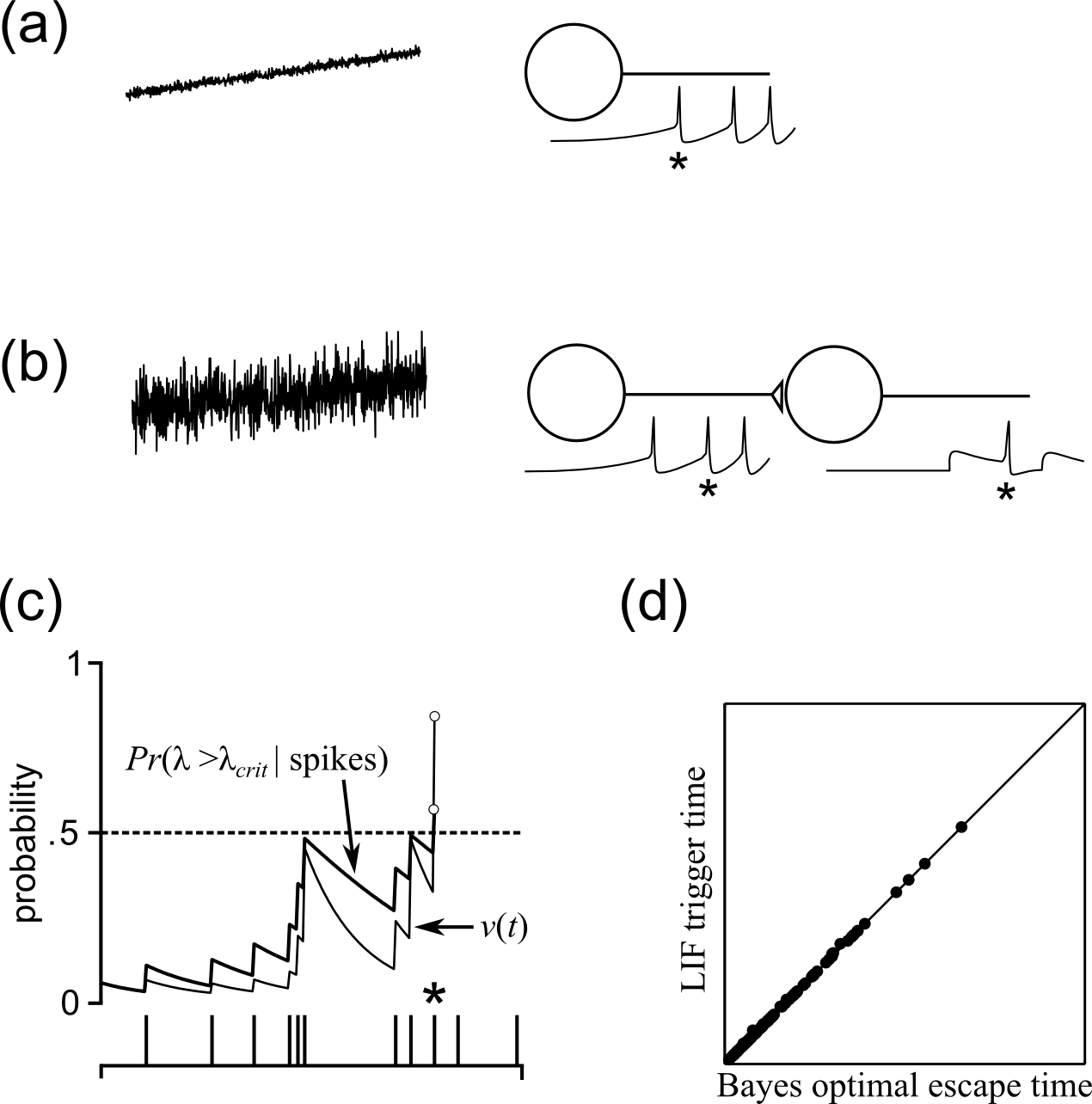}

\caption{ \label{Cap:LIFTRIGGER} (a) With high signal-to-noise levels (cf figure \ref{Cap:EDIACARAN}(c)(i)), a single neuron can trigger optimal escape behavior. (b) At moderate signal-to-noise levels, a secondary integrate-and-fire neuron can generate near-Bayes optimal escape, because the true location of the carnivore can be estimated by locally averaging the sensory firing rate (cf figure \ref{Cap:EDIACARAN}(c)(ii)).  (c) Simulation example, showing posterior probability that the carnivore is in the critical zone and the membrane potential, $v(t)$, of a secondary neuron given a sensory spike train. (d) If the critical probability is near 0.5 (the primitive condition) then the two-neuron cascade can be Bayes optimal on almost all trials.  The plot shows secondary neuron firing time vs  Bayes optimal escape time, on 100 trials.}
\end{figure}

Simulations confirm the intuition that under relevant conditions a secondary neuron that integrates the spike train and triggers at a fixed threshold can generally reduce the timing variability of escape decisions around the critical escape time.  Indeed in some regions of parameter space, secondary neuron spikes are almost always Bayes optimal. They occur precisely at the time of the sensory spike that causes the critical decision criterion to be met, on almost all trials.  

An example trial from simulations is shown in figure \ref{Cap:LIFTRIGGER}(c,d).  The simulation results are not presented in detail here because the performance of this mechanism depends quantitatively on parameters in a complex way, and these details are not required for the development of the model.  

The key observations are that when a target signal is buried in additive noise, making it effectively invisible to direct measurements, (1) threshold triggering is a kind of stochastic resonance \citep{Moss2004}, providing highly visible pulse trains that are samples from a Poisson process whose intensity depends on the target signal; and (2) the expected spike rate in these samples is the average intensity of the Poisson process \citep{Brillinger01}.  If the target signal is changing slowly relative to the spike rate, any mechanism that smooths or averages the spike train over at least a few spikes can be used to estimate the underlying target signal, and to trigger a response at a specified threshold.  Our hypothetical Ediacaran ancestors had recently acquired just such a mechanism, capable not only of smoothing a spike train by low-pass filtering, but also of triggering a spike when the average rate of the input spike train reached a specified threshold: a spiking neuron (Figure \ref{Cap:LIFTRIGGER}(a,b)).

Applying the same argument which led to the conclusion above that threshold triggers are likely to have evolved if they were possible at the onset of carnivory, it is likely that these animals would quickly have discovered that placing two of these devices in cascade provides critical threat-detectors that are effective at  lower signal-to-noise ratios than spiking neurons operating alone.  It is not necessary for these paired devices to be Bayes-optimal.  If there are inheritable developmental genetic mechanisms which would allow a subgroup of spiking sensors to migrate underneath the epithelial spiking sensors and receive inputs from them rather than directly from the environment, then they are almost certain to invade the entire population even if they offer only an arbitrarily small net benefit on average \citep{Monk2014b}. 

The two-neuron solution is Bayes-optimal in the zero noise limit, i.e. in the ancestral condition in which the single-neuron reflex is Bayes optimal. This is because if the primary neuron fires at the Bayes optimal time, then it could trigger a secondary neuron to fire at the same time.  In the ancestral condition, the first primary neuron spike triggers an escape maneuver and following spikes would be ignored.  But as noise starts to affect sensory neuron spike timing, there will be a smooth transition from zero to some high firing rate in the primary neurons, which will start firing before the predator reaches the critical state.  

For low noise levels the primary neurons remain nearly Bayes optimal, with a tendency to fire a little too early. But secondary neurons can do better, by low-pass filtering the primary spikes and triggering only when their rate exceeds a threshold indicating that the predator has reached the critical state.  At some noise level the benefits of using two neurons to trigger an escape maneuver instead of just one may exceed the costs, and then mutants that do this will almost surely invade the population.  

Two-neuron triggers would be selected not because they are Bayes optimal but because they are better than nothing. But the fact that they are not Bayes optimal means that it is possible in principle for them to be out-performed.

\section{Neurons are natural computers for inference from Poisson samples}\label{SECTION:Analog}

For a fixed value of $\lambda$, the numerator in equation \ref{eq:BayesRule} is the passive response of a first-order linear low-pass filter or leaky integrator,

\begin{equation}\label{eq:LowPassFilter}
\frac{dv(\lambda,t)}{dt} + \lambda v(\lambda,t) = 0,
\end{equation}
with initial condition 
 \begin{equation}\label{eq:IC}
v(\lambda,0) = e^{\lambda \Delta t}\left( e^{\lambda r} - 1 \right) f \left(\lambda,0 \right).  
\end{equation}
where  $f \left(\lambda,0 \right)$ is the initial prior, the probability density at $\lambda$ at time $t=0$.  

The dependent variable, $v$, in equation \ref{eq:LowPassFilter} is probability density at a point. This is a dimensionless, non-physical quantity.  In order to compute probability density with a physical mechanism it is necessary to represent it using some state variable of the mechanism.  If the mechanism uses voltage to represent probability density then $\lambda = 1/(R C)$, where $R$ is a resistance and $C$ is a capacitance. Then equation \ref{eq:LowPassFilter} can be written

\begin{equation}\label{eq:One-compartment-neuron}
C\frac{dv(\lambda,t)}{dt} + g_0 v(\lambda,t)  = 0,
\end{equation}
where $g_0 = 1/R$ is a fixed conductance.  The mechanism's time constant, which quantifies how fast it responds to a step change in input, is $\tau = 1/(R C)$. 

Equation \ref{eq:One-compartment-neuron} is the standard model of the dynamics of the membrane potential of a single-compartment neuron \citep{Hille01,Koch99}.  $C$ is its membrane capacitance and $g_0$ is its resting channel conductance. Historically this model was developed using electrophysiological studies of real neurons,  but here it has been derived as an ideal computing element for probabilistic inference from point-process observations.  A single compartment neuron exactly computes the Bayes numerator for this inference problem, when membrane potential is identified as probability density. Equation \ref{eq:One-compartment-neuron} predicts the membrane time constant for a neuron that computes probability density at a specified point.

\section{Computational role of the synapse}\label{SECTION:Synapse}

Equation \ref{eq:IC} shows that if the $k$th spike arrives at time $t_j$ then the membrane potential of a neuron with membrane time constant $\tau = 1/\lambda$ must be reset to 
 \begin{equation}\label{eq:Init}
v(\lambda,t_j^+) = e^{\lambda\Delta t}\left( e^{\lambda r} - 1 \right) f \left(\lambda,t_j^- \right),
\end{equation}
where $f \left(\lambda,t_j^- \right)$ is the prior density at $\lambda$ immediately before the spike. This ensures that the membrane potential at time $t_j^+$ immediately after the spike matches the required initial condition for computing the Bayes numerator during the following interspike interval.  

The prior density for the interval following a spike is the posterior density at the end of the preceding interval.  Suppose that there is a mechanism which continuously normalizes the membrane potential of the neuron, so that it represents posterior density rather than merely the Bayes numerator.  Then comparing the numerator in (\ref{eq:BayesRule}) to the required initial condition (\ref{eq:Init}) shows that the neuron's membrane potential must jump by a factor

 \begin{equation}\label{eq:synapticWeight}
C_s(\lambda,r) = e^{\lambda \Delta t}\left( e^{\lambda r} - 2 \right),
\end{equation}
at the spike time. A step of this magnitude is obtained by applying an impulse of magnitude $g(\lambda,r) v(\lambda, t_j^-)$ at the spike time. Adding such an impulse as an input on the right hand side of \ref{eq:One-compartment-neuron} gives

\begin{equation}\label{eq:neuron-and-synapse}
C\frac{dv(\lambda,t)}{dt} + g_0 v(\lambda,t)  = C_s(\lambda,r) v(\lambda, t) \sum{\delta(t-t_j)},
\end{equation}
where $\delta(t)$ is a unit impulse, which integrates to 1 over any interval containing 0, and to zero otherwise.  

The input must be a current (because the left hand side has dimensions of current) therefore $C_s(\lambda,r) \delta t$ is a conductance.  In this idealized form, an infinite current flows for an infinitesimal interval at the spike arrival time, instantaneously injecting a charge proportional to the membrane potential.  In a physical implementation, a finite current must flow for a finite time to inject the same amount of charge.   

A synapse causes a brief change in conductance of a neuronal membrane when a spike arrives.  This was discovered by empirical studies on neurons \citep{Hille01,Koch99}, but here it has been shown analytically that such a mechanism is necessary if a neuron is used for Bayesian inference from a Poisson process.  Equation \ref{eq:synapticWeight} predicts the synaptic conductance of a neuron that computes probability density at a specified point.

 \section{Monte Carlo integration for the Bayes Denominator}\label{sec:Denominator}

To obtain (\ref{eq:neuron-and-synapse}) it was necessary to assume that there is a mechanism for normalizing the membrane potential.  Normalization means that the posterior density must integrate to unity over all possible states.  The prey, being Bayesian, must compute the likelihood times the prior everywhere the attacker might be, in order to determine the probability density that it is at any particular location.    

Applied Bayesian inference was stalled for decades, except in simple special cases, because of the numerical difficulty of computing the kinds  of integrals that appear in the denominator of Bayes rule.  Random sampling or Monte Carlo integration techniques were developed in the 1940s \citep{Eckhardt87}, and within the last few years computers have become powerful enough to use these techniques for non-trivial Bayesian inference problems \citep{Gelman13}. 

The simplest Monte Carlo integration algorithm, called the rejection method, is to draw a random sample from a uniform distribution over the parameter space, then for each sample point (called a proposal point) draw a second uniform random number (a rejection level) between 0 and the maximum value of the integrand.  A proposal point is accepted if the rejection level is less than the value of the integrand at the proposal point. The proportion of proposal points that are accepted is the ratio of the required integral to the known volume of the space from which the proposals and rejection levels were drawn. 

For example, the numerical value of $\pi$ can be estimated by choosing random numbers uniformly between -1 and +1 in two dimensions, counting the proportion that fall within a circle of unit radius, and multiplying by 4, the area of the region in which the proposal points are scattered. By the law of large numbers this converges to $\pi$.  

If there is large uncertainty in the Bayesian posterior for an inference problem, then errors in a Monte Carlo estimate due to sampling error become negligible for small sample sizes \citep{Bolviken2001}. For example, if the area of a unit-radius circle is estimated by rejection using only 4 proposal points, then the most likely outcome, with probability exceeding $40\%$, is that 3 out of the 4 proposals will be accepted on any given trial. The area estimated from such a sample ($3.0$) may be close enough to the true value often enough to provide a competitive advantage, should you find yourself in a situation in which reproductive success depends on knowing the numerical value of $\pi$ and your competitors haven't a clue. Escape decision problems, especially in animals with limited sensory systems, are of this kind. A hugely costly decision must be made quickly on the basis of limited evidence.

\begin{figure}[h]\label{fig:reject}
\centering
\includegraphics[width = 0.5 \textwidth]{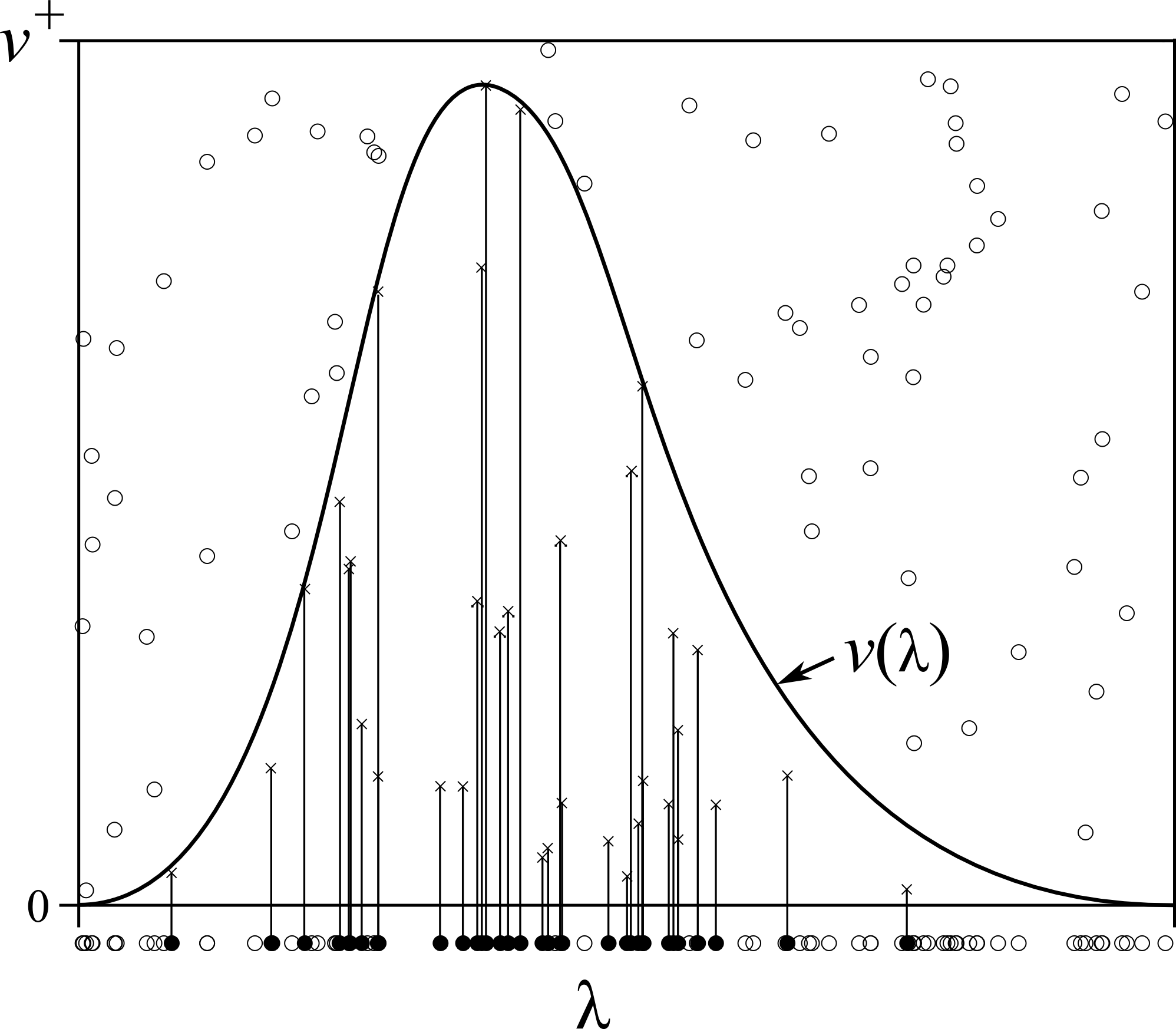}

\caption{ \label{Cap:reject} Monte Carlo integration by rejection: The area under the curve  can be estimated by drawing a proposal point, $\lambda_k$, uniformly at random in an interval containing the true parameter value, then choosing a rejection level, $u_k$, uniformly at random between 0 and a number $v^+$ larger than the maximum of the integrand. The proposal is accepted if the value of the integrand at the proposal point is above the rejection level.  The ratio of the number of accepted points (filled) to the number of proposals is the area under the curve relative to the (known) area of the enclosing rectangle.  The same result is obtained if we construct a Poisson process at each proposal point with intensity proportional to $\lambda_k$, and take a snapshot of events at an arbitrary time.  The accepted points at each instant form a random sample from the integrand treated as a probability density function.  }
\end{figure}

Suppose that there is a population of secondary neurons receiving synaptic inputs from a sensory neuron, with membrane time constants and synaptic weights given by equations \ref{eq:synapticWeight} and \ref{eq:neuron-and-synapse} corresponding to a random sample $\lambda_k, k=1,2, \dots $of proposal points over the parameter space. Suppose that each of these neurons generates a spike train that is a random sample from a Poisson process with intensity proportional to its membrane potential.  Then, by construction (because a spike has a finite duration) the probability that a neuron corresponding to a sample point $\lambda_k$ is firing at any instant is proportional to the posterior density of $\lambda$ at $\lambda_k$ at that instant.  The expected number of neurons firing at any instant is proportional to the required integral (Figure \ref{Cap:reject}). Thus if the secondary neurons fire randomly, the integral can be estimated at any time simply by counting spikes.

Rejection sampling from a uniform proposal density is rarely used except for simple example problems. This is because acceptance probability tends to be very small in high-dimensional problems and the main cost in this algorithm is in generating proposal points, so the cost scales badly with the dimensionality of the problem. However, in the spiking algorithm, each neuron represents a proposal point and each spike represents an accepted sample at that point.  The metabolic cost of neurons is dominated by the energetic cost of spiking \citep{Laughlin04,Laughlin98}, so the cost of the neural algorithm will scale predominantly with the number of  spikes, not the number of neurons.

 \section{Normalization}

If secondary neurons fire randomly with mean intensity proportional to membrane potential then the computation could be normalized by  feeding back a spatial sum of spike counts to the population, in such a way that the total depolarization is held at least roughly constant.  It doesn't matter what that constant is. Its value will determine the scaling of membrane potential, and firing intensity, to probability density.   

\begin{figure}[h]\label{fig:circuit}
\centering
\includegraphics[width = 0.5 \textwidth]{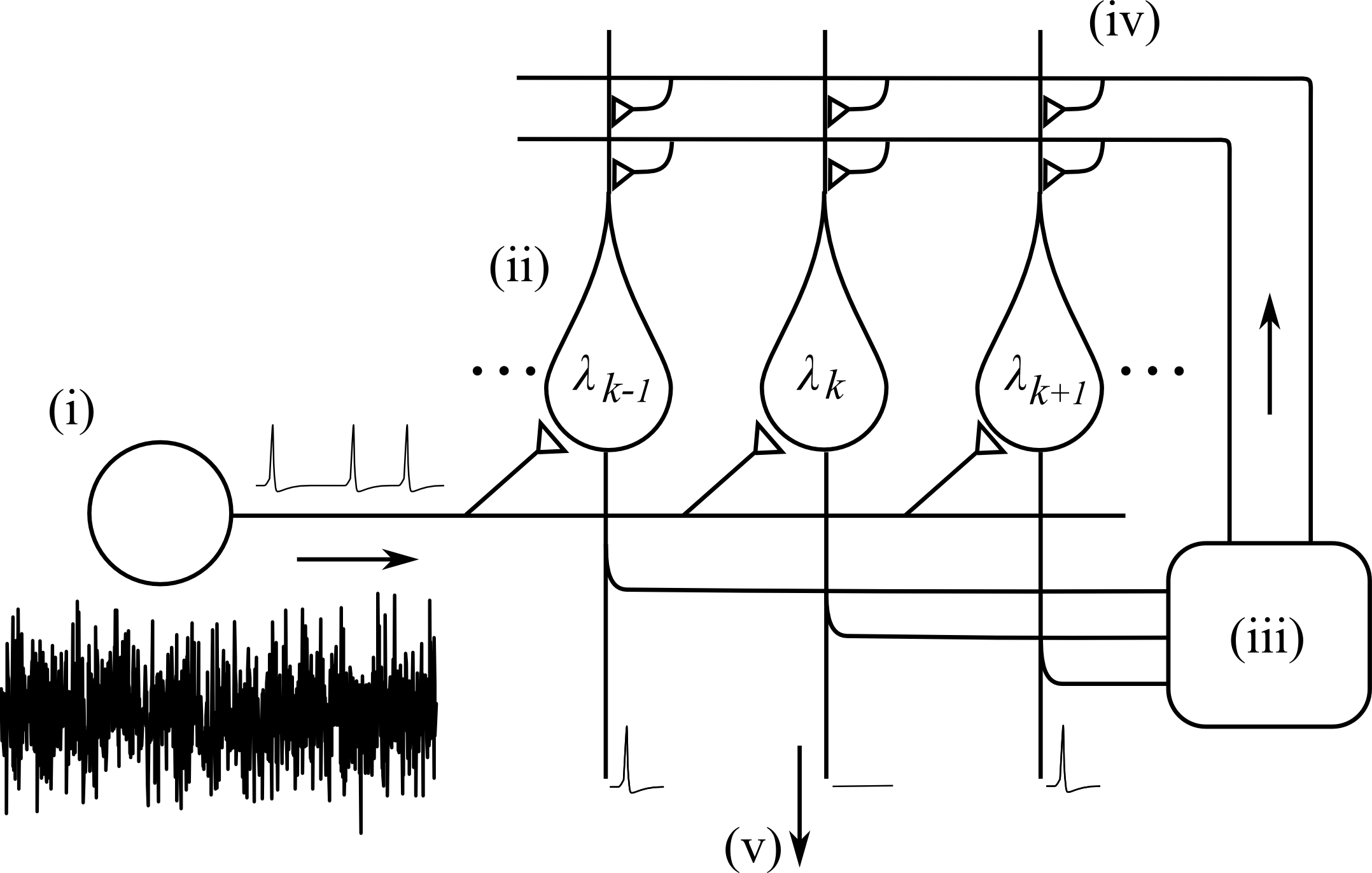} 

\caption{\label{FIGURE:Network} Neuroanatomy for Bayesian inference from a Poisson process using spiking neurons. Spikes arriving from the left are distributed to an array of observer neurons with membrane time constants $\tau_k$ corresponding to Poisson intensities $\lambda_k$.  Total spiking activity in the observer population is broadcast back on a feedback loop. This feedback signal has a multiplicative (gain control) effect on the observer neurons. It normalizes their activity and makes the spatial density of spikes in the observer population proportional to the posterior density of $\lambda$. }
\end{figure}

Normalization or stabilization of neural population activity is widespread in nervous systems \citep{Louie13,Beck2011,Eliasmith2011,Heeger92}, so much so that a recent review has referred to it as a `canonical neural computation' \citep{Carandini12}.  A number of different models have been proposed, and different mechanisms appear to be used in different systems.  In the simple case considered here it is possible to specify general anatomical and physiological characteristics that a neural normalizing mechanism must have.  A third population of neurons must sum or average activity over secondary neurons, with a short time constant, and broadcast this signal or some transformation of it back to the secondary population.  The broadcast must transmit the same signal to all secondary neurons, and have an effect on their membrane potentials equivalent to dividing each by the total activity level (Figure \ref{FIGURE:Network}).

Figure \ref{FIGURE:Network} shows the natural architecture of a neural network for inferring the posterior density of a state variable from observations of a Poisson process parameterized by that variable.  Central sensory nuclei with this general cerebellar-like architecture have been reported in invertebrate \citep{young73, young76,Farris11} and vertebrate nervous systems \citep{bell2002, bell08, Oertel14}.  They are generally found in association with sensory systems underlying perception and control of rapid movements \citep{baumann15}.  

Biophysical models can explain how spiking on parallel fibers can have the required divisive effect secondary neuron responses in cerebellar-like nuclei \citep{Nelson94}. Other models of how activity on one set of synapses can adjust sensitivity to inputs on other synapses have been proposed \citep{Burkitt03}.  Parallel fiber activation in cerebellar-like nuclei of fish has been shown to have such an effect on secondary sensory neuron activity \citep{Bastian86,Mehaffy05,Rotem07}.


\section{The neural particle filter}\label{sec:ParticleFilter}

An algorithm that draws random samples in real time from the posterior density of the state vector of a dynamical system, given a sequence of observations that are noisy functions of the state, is called a particle filter \citep{Doucet01}. Particle filters are used for Bayes-optimal nonlinear dynamical estimation, decision and control tasks. The model proposed here (Figure \ref{FIGURE:Network}(c) is a neural analog of a Bayesian particle filter. 

\begin{figure}[h]\label{fig:NPF}
\centering
\includegraphics[width = 0.75 \textwidth]{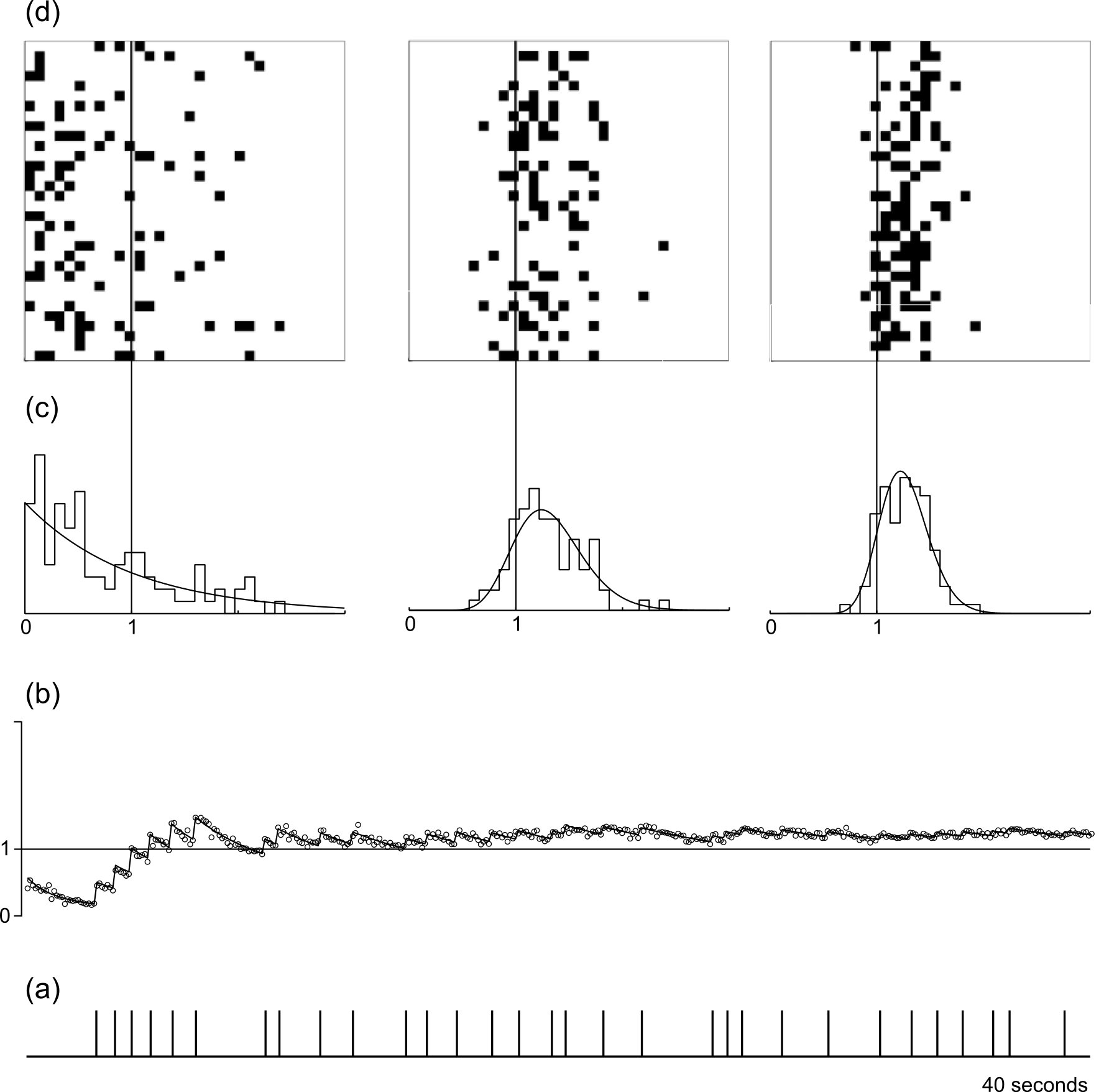} 

\caption{\label{FIGURE:ParticleFilter} Simulation of neural Bayesian particle filter. (a) Sample sensory spike train. (b) Exact Bayesian posterior median (solid) and neural Bayesian posterior median (circles) at sample times. (c) Spatial histograms of spike locations in three snapshots, mapped in $\lambda$-space, overlaid by the numerically computed posterior density of $\lambda$ at the corresponding times. (d) 2-D spatial maps of spiking activity at the snapshot times. Columns in the 2D map align with bins in the spatial histograms.}
\end{figure}

Figure \ref{FIGURE:ParticleFilter} shows output from a simulation of a neural particle filter, written in MATLAB. There's a signal source  at a unit distance causing sensory neurons to generate spikes that are samples from a Poisson distribution with mean rate 1/s.  The filter has 1024 secondary neurons, each described by equation \ref{eq:neuron-and-synapse}, with membrane time constants spaced uniformly to represent signal source locations up to 3 times more distant than the true source location. The simulation time step length is 1.0ms.  Each secondary neuron fires randomly in each timestep with probability proportional to its membrane potential.  All secondary neuron membrane potentials are divided in proportion to the total number of spikes in the secondary population on the previous timestep. This division is scaled so that the expected number of spikes in the secondary population is 128, $1/8$ of the population size.  

The lower trace (a) in this figure shows a  Poisson spike train lasting 40s, with Poisson intensity $\lambda^*=1$ spike per second. It is left-censored with a refractory period of 500msec.  This input is distributed to all secondary neurons, which respond according to equation \ref{eq:neuron-and-synapse}.

The solid line in the next trace (b) shows the median of the posterior density of  $\lambda$ given the spike train, computed numerically from Bayes rule (\ref{eq:BayesRule}) with an exponential initial prior at $t=0$. This shows the true posterior median, within numerical error.  The posterior median jumps to higher values of $\lambda$ at spike times and drifts back towards zero during interspike intervals. Open circles in (b) represent the median secondary spike at each timestep. The median spike is the spike such that half of the spikes in the population at that instant are occurring in neurons that map points further away and half are occurring in neurons that map closer points, than the neuron in which this spike is occurring.  It is the sample point that divides the sample into two regions of equal probability mass above and below it.  

The three graphs in \ref{FIGURE:ParticleFilter}(c) are spatial histograms of spike locations in the secondary population, with $\lambda$ divided into 32 bins of equal width, in `snapshots' of secondary neuron activity taken at $t=0$, $t=20s$ and at $t=40s$.  The true parameter value, $\lambda=1$, is shown in each plot by a vertical line.  The horizontal axes in these plots are values of $\lambda$, and the vertical axes represent posterior density. The width of a column represents a region in the parameter space and the height of a column represents the number of neurons corresponding to sample points in that region that were firing when the snapshot was taken.  The smooth curves show the true posterior densities at the snapshot times.  The initial spike histogram is a random sample from the exponential initial prior density.  As spikes arrive the true posterior density becomes increasingly Normal and narrower and the neural particle filter infers it accurately.  

The three graphs in \ref{FIGURE:ParticleFilter}(d) illustrate the `native' representation of probability density in a neural particle filter. In these plots, the 1024 proposal points corresponding to secondary neuron time constants are laid out consecutively in 32 columns.  These can be interpreted as maps of secondary neuron time constants, or as maps of $\lambda$-space, or as maps of distance to the predator.  Real space is distorted in the map by the inverse cubic transformation between distance and Poisson intensity.  Dark squares show active neurons at each of the three snapshot times.  Columns of the arrays in (d) match the histogram bins in (c).  The two-dimensionality of the maps here is for visualization purposes; the map is functionally one-dimensional in parameter space, and in real space.

In the displayed trial the neural particle filter overestimates the value of $\lambda^*$ at the end of the simulation period, but this is an accurate estimate of the true posterior density.  Most of the posterior probability mass is above the true value of $\lambda$ in this trial because this particular sample has more spikes than expected.  In reality almost every spike train will have more or fewer spikes than the expected number given the true parameter value, because spiking is a stochastic process. In this trial (as in almost every trial) the observer has an incorrect belief about the location of the signal source, but Bayes optimality requires taking this belief seriously. 
 
The neural particle filter circuit was derived under the assumption that the Poisson parameter is fixed. The carnivore is not moving, and the expected firing rate of the sensory neuron is constant. Under these conditions posterior density estimates are asymptotically exact for large numbers of neurons, because the Monte Carlo estimate converges to the true denominator and then the neural equations are exactly Bayes rule for inference from a Poisson process. A limiting-case carnivore that does not move is not very interesting or threatening, but  the earliest carnivores must have moved very slowly.  Carnivory invaded Ediacaran ecosystems starting from the limiting case of no movement.

\section{Bayes optimal escape with small networks}
\begin{figure}[h]\label{fig:EscapeTiming}
\centering
\includegraphics[width = 0.75 \textwidth]{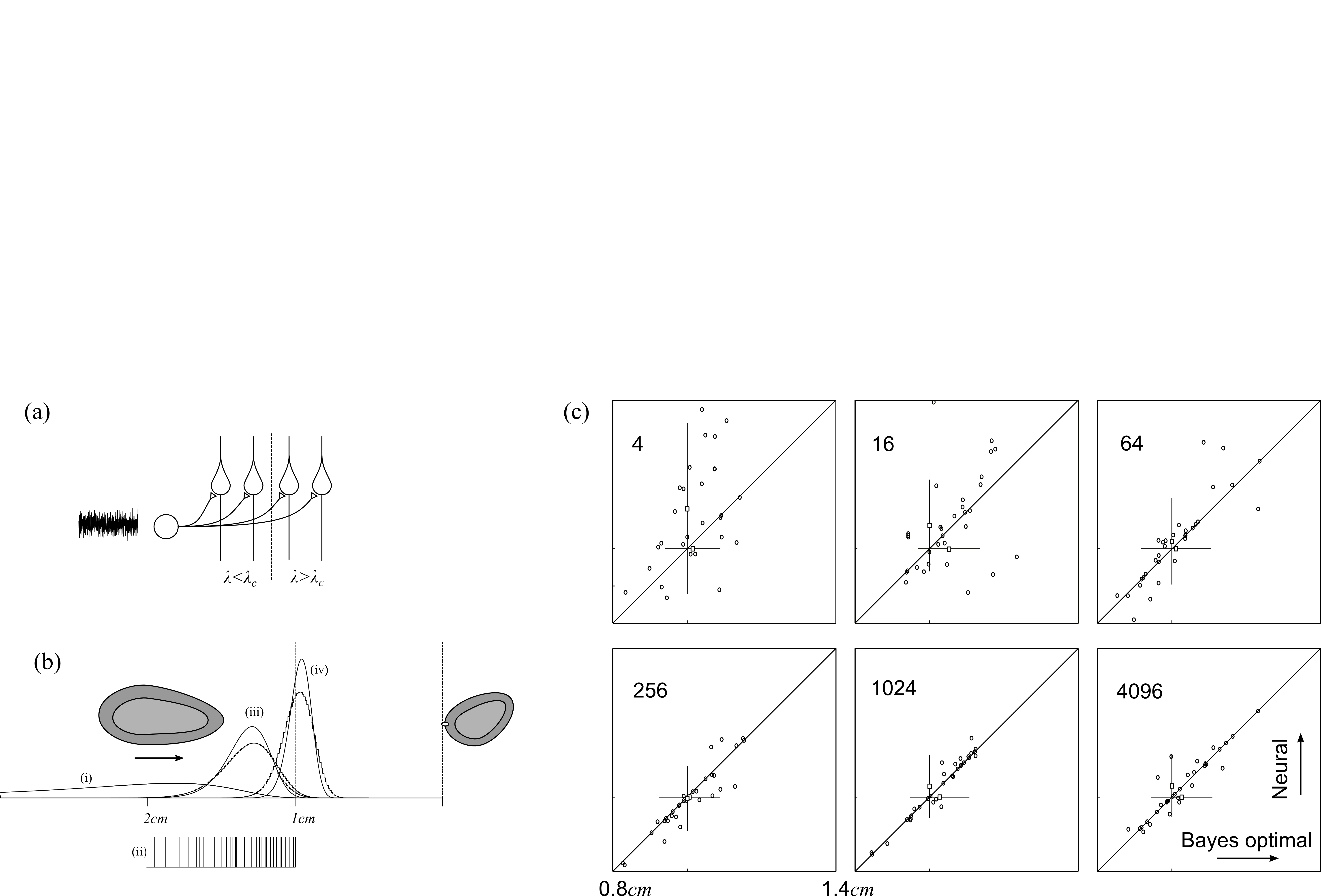} 

\caption{\label{FIGURE:EscapeTiming} Escape timing decisions by small networks.  (a) A 2-neuron trigger (Figure \ref{Cap:LIFTRIGGER}) can be modified by duplicating secondary neurons.  Adding a feedback loop converts the array of secondary neurons into a neural Bayesian particle filter.  (b)  A carnivore approaches a grazer, causing sensory neuron spikes to occur at a faster rate. The grazer escapes when when the posterior median distance to the carnivore, computed via a neural particle filter, reaches the critical distance.  (b) Distance to the carnivore at the escape time plotted against the distance at the Bayes optimal escape time. Grazer escape decisions approach Bayes optimality for modest neural population sizes. Details in text.  }
\end{figure}

A neural particle filter (\ref{FIGURE:Network}) could evolve from a simple two-neuron precursor (Figure \ref{Cap:LIFTRIGGER}(b)) via a sequence of small modifications.  If there were several secondary neurons instead of just one, with thresholds scattered to trigger at different points near the critical distance (Figure \ref{FIGURE:EscapeTiming}(a)), this could plausibly provide utility by making it possible, for example, to take preparatory action as a carnivore approaches but before it reaches the critical danger zone. This secondary array can become a Bayesian particle filter, simply by adding tertiary neurons that spatially sum activity over the array and feed it back as a normalizing signal.  Tertiary neurons that sum activity over a part of the array could compute the posterior probability that the carnivore is in the corresponding range of locations. In particular, they could estimate the posterior probability that the carnivore is closer than the critical distance.

A particle filter is more complicated to build and more expensive to operate than simply using secondary neurons to compute running means of sensory neuron firing rates, because it needs a large population of secondary neurons and an additional population of tertiary neurons.  On the other hand, a particle filter can compute the required posterior probability with arbitrary accuracy if it has a sufficient number of neurons.  Secondary neurons acting as leaky integrators work well if the sensory signal-to-noise ratio is high at the critical distance, because in that case the posterior density will be narrow (approaching a Gaussian) when the carnivore approaches that point, and there is a high probability that the posterior mean will be near the true location of the carnivore. The mean of the posterior density is the mean firing rate, and so triggering when the mean afferent firing rate crosses the threshold rate corresponding to the critical distance will be close to  Bayes optimal.  

At low SNR, the posterior density is broad and skewed when the carnivore is near the critical distance, and the posterior mean may be well outside the critical zone when more than $50\%$ of the posterior probability mass is already inside it.  Because there is a high expected cost for even a small probability that the carnivore is very close, the critical probability, $p_{crit}$, will become lower than 0.5 at lower SNR.  Leaky integrator neurons that compute average firing rates can no longer be used to make Bayes optimal decisions, but they could reorganize themselves to form a neural particle filter and cooperate to compute Bayesian posterior probabilities. 

The performance of neural particle filters with small numbers of neurons has been examined by simulating escape from a slowly advancing predator (Figure \ref{FIGURE:EscapeTiming}).  The parameters of the model are arbitrary, but specific numbers have been chosen because this makes the simulation results easier to understand.  A carnivore approaches a grazer at 1mm per second, starting at 2cm (Figure \ref{FIGURE:EscapeTiming}(b)). The critical distance is 1cm.  The sensory neuron (i) fires with intensity $8/x^3$ when the distance between carnivore and grazer is $x$.  The initial prior density is exponential in $\lambda$ with expected intensity 1 spike per second. This transforms to a prior assumption (ii) that it  is initially very unlikely that there is a carnivore closer than the critical distance, but there is likely to be one within a few times that distance.  There's also an implicit prior assumption that there's only one carnivore in the neighborhood.  

The true posterior density of $\lambda$ given the spike train up to time $t$ is computed by numerical sequential Bayes based on the true model, with 1ms update interval.  This entails applying Bayes rule (\ref{eq:BayesRule}) and sliding the posterior density of carnivore location uniformly towards the grazer at 1mm per second. The grazer is assumed to not (yet) incorporate an accurate model of carnivore movements into its inferences about carnivore location. Instead, the neural particle filter's posterior density estimate is broadened by a small amount on each timestep, adding uncertainty to allow for the possibility that the source may be moving.  

Neural particle filters with 4, 16, 64, 256, 1024 and 4096 secondary neurons were simulated. Secondary neuron  time constants were spread uniformly over values corresponding to carnivore locations between half and three times the critical distance.  The carnivore's approach was simulated 32 times for each neural population size. The Bayes optimal escape time was computed as the timestep in which the true posterior probability that the carnivore was inside the critical zone first exceeded 0.5.  The grazer's escape time was computed as the first sensory spike time after which the number of spikes in the secondary population in neurons representing locations closer than the critical time exceeded half the expected number of spikes in the population.  

For example, with 4 secondary neurons, just two neurons map to distances beyond the critical distance and the other two map to distances closer than the critical distance. The feedback gain is adjusted so that on average two secondary neurons fire at any one time. An escape maneuver is generated if at least one of the neurons representing points inside the critical zone fires after a sensory spike.  

Figure \ref{FIGURE:EscapeTiming}(b) shows the true posterior and the posterior computed by a neural particle filter, before the carnivore reaches the critical point (iii), and when both the true and the neural estimate of the posterior probability that the carnivore is in the danger zone exceeds 0.5 (iv).  The posterior density estimated by the neural particle filter is flatter than the true posterior and has a slightly smaller area.  The relative flatness is due to the neural observer's uncertainty about target movement, compared to the ideal Bayesian observer's inferences based the true prior belief that the carnivore is moving at a specific constant speed.  The difference in areas is because the neural filter is only approximately normalized, using a random sample spike count, compared to the exact normalization of the ideal Bayesian.  

Figure \ref{FIGURE:EscapeTiming}(c) shows comparisons between the distance to the carnivore at the Bayes optimal escape time versus the distance at the escape time computed by a neural particle filter, for the different secondary neuron population sizes.  The horizontal and vertical axes each run from 0.8cm to 1.4cm, which includes all of the data, except for one trial of the 4-neuron filter in which the distance at the Bayes optimal escape time was less than 1cm but the grazer's escape was triggered when the carnivore was near 2cm.  Circles representing individual trials lie along the diagonal if the grazer's decisions are Bayes optimal.  The open squares with bars represent mean and standard deviation of the distance to the carnivore at the Bayes optimal  (horizontal) and grazer  escape times (vertical). 

These plots show that a neural particle filter with just a handful of neurons can compute posterior probabilities roughly, and decisions made using a neural particle filter become increasingly closer to Bayes optimal as the size of the secondary neuron population increases.  

A trigger mechanism for escape maneuver timing does not need to be Bayes optimal in order to be selected by evolution. It only needs to be better than available alternatives.  A neural particle filter network that reduced uncertainty in primary afferent data, and reduced losses due to poor escape decision timing, could be reached by a simple sequence of modifications of existing mechanisms in Ediacaran animals. The decision network could evolve to make decisions that were arbitrarily close to Bayes optimal, simply by enlarging its core component -- the neural Bayesian particle filter.

\section{Discussion}

The evolutionary scenario presented in this paper, in which the first carnivore is portrayed as a placozoan-like motile animal attacking algal mat grazers that had a similar morphology, is only one possibility. Evidence of carnivory appears in the fossil record starting at about 550Ma, and escalates slowly until the relatively sudden appearance of most modern animal phyla  in the Cambrian explosion twenty million years later  \citep{ErwinBook2013, Marshall06}.   But carnivory is likely to have preceded any of these expensive innovations \citep{Monk2015b}.   By the middle of the Ediacaran period there were many species of animals, and the first carnivore among them was probably a soft blob that has left no trace. It may have resembled a creeping fungus as much as any modern animal.  Neuron evolution could have been stalled by environmental constraints, such as low oxygen levels, or biotic constraints, such as a lack of patterning genes, for many millions of years after carnivores first appeared. After environmental and developmental conditions made them possible, and ecological conditions them necessary, neurons could have evolved independently many times.  Molecular evidence indicates that neurons and nervous systems did evolve at least twice during the Ediacaran Period, in bilaterians and in ctenophores \citep{moroz2015,Halanych2015}.  

Many uncertainties about the detailed historical path of neuron evolution remain, but these do not affect the principle or the conclusions of the present model.  It demonstrates that Bayesian brains could have bootstrapped themselves out of ciliated epithelial tissues in a remarkably simple way, given the right incentives under the right conditions. Suitable incentives and conditions first occurred on Earth during the Ediacaran Period.  

The model explains fundamental properties of neurons and nervous systems.  It predicts the existence of spiking neurons. According to the model, spiking neurons evolved at the onset of carnivory to detect cues from an approaching carnivore and signal the precise, utility-maximizing escape time.  

The model predicts that sensory spike trains can be modeled as observations from refractory-censored Poisson processes parameterized by relevant state variables.   According to the model, this occurred soon after neurons first evolved, because of selection pressure for increased sensitivity.   Spiking neurons were transformed from detectors of particular world states into a mechanism for extracting information about world states from weak signals buried in noise, a form of stochastic resonance. This transformation does not require any additional explanation. It's simply what happens when the threshold level of a threshold trigger drops below the noise level of its input, or the input signal drops below the noise level of the triggering mechanism.

Refractory censoring of sensory spike trains is predicted because spikes within some fixed interval after a preceding spike must ignored for real time inference from a Poisson process.  Such spikes would be expensive but useless, and so a mechanism to suppress them should evolve.  Because equation \ref{eq:BayesRule} is exact for arbitrarily long refractory periods, the duration of interspike intervals in sensory spike trains could be dominated by the refractory mechanism and therefore be more regular than expected for a Poisson process.   If the model is correct then interspike interval distributions in primary afferent spike train data used for state estimation tasks (for example, electrosensory or vestibular) should be exponential convolved with the latency distribution of a censoring process.  This is a simple, testable prediction of the model.  

The model predicts that primary afferent neuron axons will diverge onto arrays of secondary neurons, and that there will be a systematic relationship between membrane time constants, synaptic input weights and the `preferred states' or state-space receptive fields of secondary neurons.  This prediction is testable in principle, but the quasi-static, one-dimensional inference problem used to derive the model here is probably simpler than any problem faced by an Ediacaran, and almost certainly much simpler than any problem faced by any animal today.  Testing this prediction will require extending the model to deal with a realistic inference and decision problem in an extant animal.

The model predicts that secondary neuron arrays will have mechanisms that stabilize their total activity levels.  This prediction is testable statistically, independent of mechanism, by showing that when sensory inputs fluctuate, increased firing rates in some secondary neurons are accompanied by decreases in others, such that the mean firing rate in a random sample from the secondary population remains roughly constant.  

In the neural particle filter  model described here there is only one sensory modality and a normalizing signal can only be constructed by feedback from the secondary population. Modern nervous systems have access to multiple sensory cues as well as to efference copy signals that can combine with feedback to generate a predictive normalizing signal.  Neurons transmitting such a signal should appear under natural conditions to respond to the same modality as the secondary neurons, but experimental manipulation and neuroanatomy should show that normalizing signals are constructed not solely by feedback from secondary neurons, but also from other sources that predict fluctuations in sensory inputs, such as efference copy of motor commands. These neurons should have broad receptive fields in their responses to sensory stimuli, because the normalizing signal represents an average over a population of secondary neurons that respond differentially depending on the location of the stimulus source. They should have  low sensitivity to natural patterns of sensory input because when normalization is accurate the average level of secondary activity, and therefore the normalizing signal, is nearly constant.  Stimulation of the descending normalizing pathway should affect the sensitivity of secondary neurons to primary afferent inputs. 

The earliest animals with spiking neurons may have used a number of cues singly or in combination to trigger attack or withdrawal reflexes when they  made contact with other animals.  When escape was triggered before contact, there must have been selection pressure not only to determine distance but to recognize or classify approaching animals, using whatever sensory cues were available.  Bayes optimality in this situation requires computing posterior probabilities for discrete causes. Particle filters are not suited to this task, but it can be done using for example message passing in a belief network \citep{BayesNetworks2003}. Deneve has developed a model of Bayesian inference and decision-making by spiking neurons, using spikes to transmit messages in a Bayesian belief network \citep{Deneve08,deneve08b}.  This shows that it is ultimately possible for neurons to do this.   Under the fundamental assumption of the present model, that Bayesian inference is necessary if it is possible in the face of mortal danger, the earliest nervous systems probably used multiple sensory cues, and developed inference mechanisms for continuous state estimation and discrete classification tasks in a coordinated way from the start.  

When animals first made contact, chemosensing would have the considerable advantage as a withdrawal/attack trigger cue because an animal could be identified by tasting it.  At a distance, odorants provide valuable information about identity but increasingly poor information about distance or direction.  Odorants could have played a role in the evolution of mechanisms for Bayesian classification analogous to the hypothetical role of electric fields in the evolution of mechanisms for state estimation. They were available as an unambiguous cue at the start, and at close range posterior probabilities for source identity could still be computed without using other sense data.  This would seem to make it relatively easy to maintain Bayes optimality as the inference problem got harder.  Although no proposal has been made here about how this could happen, the implication is that at some time preceding the Cambrian explosion, nervous systems in motile animals consisted of a multisensory Bayesian dynamical state estimator elaborated from a circuit for electroreception, and a multisensory Bayesian classifier elaborated from a circuit for olfaction. These circuits would collaborate (because movement patterns provide information about identity, and vice-versa) and would project to decision-making structures that projected to muscles. 

Vertebrate nervous systems, with their relatively sophisticated capabilities for causal inference and agile motor control, have traditionally been regarded as more advanced than those of invertebrates, with the implication that they evolved from  primitive nervous systems resembling the simpler nervous systems of modern invertebrates.  But that is not the pattern indicated by the fossil record or by comparative neurobiology, and not what is expected under the current model.  Chordates, the phylum containing vertebrates, including us, appear in the fossil record during the Cambrian explosion, simultaneously with almost all other modern animal phyla.  Being clever and agile appears to have been an option from the beginning. 

There are many possible evolutionary responses to predation threats, but it isn't cost-effective for an oyster to ponder the identity and intentions of nearby animals in great detail, nor does closing its shell call for very much  dexterity.  Nervous systems of modern animal phyla -- like other aspects of animal morphology at the phylum level  -- appear to have radiated from a poorly differentiated common ancestor, rather than via a series of steps linking different morphologies.  The model presented here suggests that Bayesian inference with Bayes optimal decision-making might be a primitive character among animals with nervous systems.  Small, simple invertebrate nervous systems in modern animals should be capable of Bayesian inference, unless it has been secondarily lost in a niche that does not require it.

The sea anemone \emph{Anthopleura aureoradiata} is a soft-bodied sessile suspension feeder with a fast withdrawal reflex triggered by mechanical disturbances. The timing and duration of the withdrawal reflex are correlated with environmental and stimulus parameters in a way that minimizes probable losses. Costs and benefits were quantified by the experimenters' observations, not those of the sea anemone, but the result is consistent with Bayes optimal decision making by the anemone \citep{Haag14}.  The tube-dwelling marine polychaete, \emph{Serpula vermicularis}, has a similar suspension-feeding lifestyle with very fast withdrawal when disturbed, but with a more elaborate battery of sensors.  Its withdrawal reflex decisions are similarly consistent with a Bayes optimal model \citep{Dill1997}.  These modern animals occupy niches resembling those of Ediacaran animals in which nervous systems first evolved.  According to the model derived in this paper, withdrawal reflexes in these animals ought to be Bayes optimal. Their nervous systems should contain a network resembling the circuit shown in figure \ref{FIGURE:Network},  intervening between sensors that measure threats and motorneurons that drive the withdrawal reflex.   Cerebellar-like circuits, whose function is presently unknown, have been reported in polychaetes and other invertebrates that make rapid movements  \citep{baumann15,Farris11,young73}.

Cerebellar-like brainstem electrosensory nuclei of sharks and certain other fish have the anatomical and physiological characteristics predicted by the neural particle filter model \citep{bell08}.  In the ancestral state, electroreception had the special feature that Bayesian inference using electric fields could be bootstrapped from a simple decision mechanism and a one-dimensional inference problem.  This is not true for other senses, and it is no longer true for electroreception, because electric fields around modern animals in water are functions of the time-varying configuration of the body \citep{bodznick99}.  In agile, articulated animals, Bayesian perception is impossible without taking into account the configuration and movement of the body. Our sense organs are located on a moving platform, and we actively explore the world rather than listening passively and waiting for something to attack us.  In animals that chose agility as a predation strategy or  a response to it, there will have been early selection pressure  to estimate body state from inertial (vestibular) and proprioceptive sense data by elaborating the electrosensory filter. This would allow state estimation using multiple cues on a moving platform.  Comparative neuroanatomy is consistent with the view that the cerebellum evolved by elaboration of electrosensory brainstem nuclei, with the vestibulo-cerebellum as an intermediate stage \citep{monty2010}.

The model, analysis and simulations in this paper suggest that nervous systems can generally be understood as an elaboration of a `tri-neuronal reflex arc' consisting of (a) spiking sensory neurons providing noisy observations of world and body states (b) populations of central neurons computing posterior probability densities of states and posterior probabilities of causes and (c) decision neurons that generate actions conditioned on these probabilities.  Probability density is represented centrally by spatial density of spikes, and the probability of an event or cause is represented by the instantaneous spike count in a sub-population of neurons.  The results show that using Bayesian models to analyze spike-based computation in simple sensory-motor pathways of vertebrate and invertebrate nervous systems, constrained by models of the ecological problems that they evolved to solve, may provide a framework for understanding the structure and function of all nervous systems, including our own.

\normalfont

\subsection*{Acknowledgments} Supported by PBRF and a University of Otago Study Leave grant. Thanks to Andre van Schaik for finding and correcting an error in equation (\ref{eq:synapticWeight}), to Travis Monk for useful discussions on neuralian evolution and quantitative methods, and to Helen Davies for assisting with manuscript preparation.

\pagebreak
\bibliographystyle{apa}
\bibliography{OI3.bbl}


\end{document}